\documentclass[aps,prb,twocolumn,superscriptaddress,showpacs,showkeys,floatfix]{revtex4}

\usepackage{graphicx,color}
\usepackage{multirow,slashbox}

\graphicspath{{figs/}}
\begin{document}
\title{Interlayer Breathing and Shear Modes in Few-Layer Black Phosphorus}
\author{Jin-Wu Jiang}
    \altaffiliation{Corresponding author: jwjiang5918@hotmail.com}
    \affiliation{Shanghai Institute of Applied Mathematics and Mechanics, Shanghai Key Laboratory of Mechanics in Energy Engineering, Shanghai University, Shanghai 200072, People's Republic of China}
\author{Bing-Shen~Wang}
    \affiliation{State Key Laboratory of Semiconductor Superlattice and Microstructure and Institute of Semiconductor, Chinese Academy of Sciences, Beijing 100083, China}
\author{Harold S. Park}
    \affiliation{Department of Mechanical Engineering, Boston University, Boston, Massachusetts 02215, USA}

\date{\today}
\begin{abstract}
The interlayer breathing and shear modes in few-layer black phosphorus are investigated for their symmetry and lattice dynamical properties. The symmetry groups for the even-layer and odd-layer few-layer black phosphorus are utilized to determine the irreducible representation and the infrared and Raman activity for the interlayer modes. The valence force field model is applied to calculate the eigenvectorw and frequencies for the interlayer breathing and shear modes, which are explained using the atomic chain model. The anisotropic puckered configuration for black phosphorus leads to a highly anisotropic frequency for the two interlayer shear modes. More specifically, the frequency for the shear mode in the direction perpendicular to the pucker is less than half of the shear mode in the direction parallel with the pucker.  We also report a set of interlayer modes having the same frequency for all few-layer black phosphorus with layer number $N=3i$ with integer $i$, because of their collective vibrational displacements. The optical activity of the collective modes supports possible experimental identification for these modes.

\end{abstract}

\maketitle
\pagebreak

\section{Introduction}

Heterostructures are a sequential stacking of two different two-dimensional (2D) layered materials  \cite{NetoAHC2011rpp}, which are coupled together via interlayer van der Waals interactions.  Characterization of the interlayer coupling in the heterostructures can be done using a lattice dynamical analysis.  Specifically, the interlayer breathing (B) mode and shear (C) mode directly represent the interlayer coupling properties in the layered materials. The frequency for the B mode is depending on the number of layers, so this mode can be used to determine the number of layers of the heterostructure, while the C mode gives insight into the friction between two neighboring 2D layers.

The C mode in few-layer graphene was examined experimentally by Tan et al. in 2012.\cite{TanPH} The frequency for the highest-frequency C mode depends on the layer number ($N$) as $\sqrt{1+\cos (\pi/N)}$, which was explained by the chain model, and is around 30~{cm$^{-1}$} in few-layer graphene. Due to its low frequency, the C mode can be excited easily, so it is sensitive to the near-Dirac point quasi-particles.\cite{TanPH}  In particular, the C mode is easily excited during cross-plane thermal transport in the layered materials, due to its low frequency. The scattering between the C mode and the acoustic modes may play an important role for the cross-plane thermal transport in the layered materials. More recently, two experiments found that the signal of the C mode can be enhanced by folding the graphene layers.\cite{TanPH2014prb,CongC2014nc} The B mode in few-layer graphene has also been studied by several groups.\cite{JiangJW2008prb,LuiCH}

As another important 2D layered material, few-layer MoS$_{2}$ has also attracted significant attention for its interlayer modes. Several experiments have measured the $N$ dependence for the frequency of the interlayer B mode and C mode in few-layer MoS$_{2}$.\cite{PlechingerG2012apl,ZengHL,ZhaoYY,ZhangXprb2013} The frequency of the interlayer B mode decreases with increasing layer number, while the C mode exhibits the opposite behavior.

Few-layer black phosphorus (FLBP) is another emerging 2D layered material with that shows an $N$-dependent band gap.\cite{LiuH2014,DuY2010jap}  However, few works have been performed for the phonon modes in black phosphorus. The phonon dispersion for bulk black phosphorus was measured\cite{FujiiY1982ssc,YamadaY1984prb} in 1980s. The experiment was explained by Kaneta et al. using the valence force field model (VFFM)\cite{KanetaC1982ssc,KanetaC1986jpsj} or the adiabatic bond charge model.\cite{KanetaC1986jpsj2}  While the layer number dependence for the interlayer B mode and C mode in the FLBP has not been studied to-date, an important and interesting effect to quantity is that of the intrinsically puckered BP geometry on the interlayer modes.  We thus analyze the lattice dynamics properties for the interlayer B mode and C mode in the FLBP.

In this paper, we study the symmetry and the lattice dynamical properties for the interlayer B mode and C mode in FLBP. The symmetry groups for the FLBP with even or odd layer numbers are compared. Using these symmetry groups, we analyze the symmetry for the interlayer B and C modes, including their irreducible representations and their infrared (IR) and Raman activity. The VFFM is utilized to compute the eigenvectors and frequencies for the interlayer B and C modes, while the calculated results are explained by the linear chain model. As a result of the intrinsic geometric anisotropy in the puckered configuration of BP, the two interlayer C modes have very different frequencies. Furthermore, we present a set of collective interlayer modes in FLBP with layer number $N=3i$ with integer $i$. The frequencies for these collective modes are independent of the layer number and these modes are optically active, so they should be experimentally measurable. 

\begin{figure}[tb]
  \begin{center}
    \scalebox{0.7}[0.7]{\includegraphics[width=8.0cm]{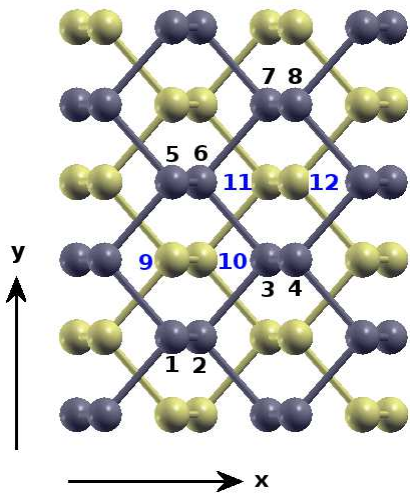}}
  \end{center}
  \caption{(Color online) Top view of bulk BP.  The two BP layers are displayed by different colors. The x-direction is perpendicular to the pucker, and the y-direction is parallel with the pucker.}
  \label{fig_cfg}
\end{figure}

\begin{table}
\caption{VFFM parameters.  The original parameters from Ref~\onlinecite{KanetaC1986jpsj} are listed in the second line. The third line lists optimized parameters used in the present work. All parameters are in the unit of eV\AA$^{-2}$. The corresponding potential is $V=\frac{1}{2}\alpha\left[\left(\vec{u}_{i}-\vec{u}_{j}\right)\cdot\hat{e}_{ij}\right]^{2}$, where $\vec{u}_{j}$ is the displacement for atom j and $\hat{e}_{ij}$ is the unit vector from atom i to atom j.}
\label{tab_vffm}
\begin{tabular}{@{\extracolsep{\fill}}|c|c|c|c|}
\hline 
 & $\alpha_{1}$ & $\alpha_{2}$ & $\alpha_{3}$\tabularnewline
\hline 
Ref~\onlinecite{KanetaC1986jpsj} & 0.321 & -0.01 & 0.015\tabularnewline
\hline 
present work & 0.281 & -0.067 & 0.03\tabularnewline
\hline 
\end{tabular}
\end{table}

\begin{table}
\caption{Frequency (in cm$^{-1}$) for the B mode, C$_{x}$ mode, and C$_{y}$ mode in bulk BP. Theoretical results from Ref~\onlinecite{KanetaC1986jpsj} (3rd line) and the present work (4th line) are compared with experiments (2nd line).  The values in parentheses (3rd line and 4th line) are the relative difference between the theoretical prediction and the experiment.}
\label{tab_frequency}
\begin{tabular}{@{\extracolsep{\fill}}|c|c|c|c|}
\hline 
 & B & C$_{x}$ & C$_{y}$\tabularnewline
\hline 
exp & 87.1 & 19.4 & 51.6\tabularnewline
\hline 
Ref~\onlinecite{KanetaC1986jpsj} & 92.8 ($\uparrow6.5\%$ ) & 21.1 ($\uparrow8.8\%$) & 53.5 ($\uparrow3.7\%$)\tabularnewline
\hline 
present work & 87.1 ($0\%$) & 20.0 ($\uparrow3.1\%$) & 51.7 ($\uparrow0.2\%$)\tabularnewline
\hline 
\end{tabular}
\end{table}

\begin{figure}[tb]
  \begin{center}
    \scalebox{1.05}[1.05]{\includegraphics[width=8.0cm]{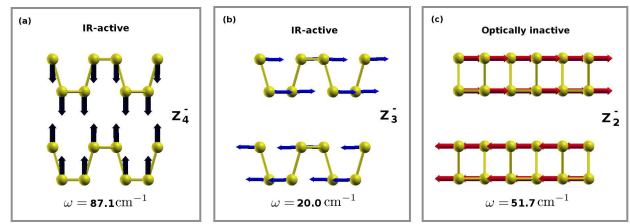}}
  \end{center}
  \caption{(Color online) Eigenvectors for the three interlayer modes in bulk BP. (a) B mode. (b) C$_x$ mode. (c) C$_y$ mode. The arrow on top of each atom represents the vibrational amplitude of the eigenvector.}
  \label{fig_bulk_all}
\end{figure}

\section{Interaction Potential}

The intralayer interaction is described by a recently developed Stillinger-Weber potential.\cite{JiangJW2014bpsw} We apply the VFFM for the interlayer coupling between two adjacent BP layers.\cite{KanetaC1986jpsj} This VFFM contains the following bond stretching interaction,
\begin{eqnarray}
V & = & \frac{1}{2}\alpha\left[\left(\vec{u}_{i}-\vec{u}_{j}\right)\cdot\hat{e}_{ij}\right]^{2},
\end{eqnarray}
where $\vec{u}_{j}$ is the displacement for atom j, $\hat{e}_{ij}$ is the unit vector from atom $i$ to atom $j$, $\alpha=$ $\alpha_1$, $\alpha_2$, and $\alpha_3$ are the parameters for the first-, second-, and third-nearest-neighbor interlayer interactions, respectively. Fig.~\ref{fig_cfg} shows the configuration for bulk BP. The interlayer first-nearest-neighbor distance is the distance between atoms 9 and 2, i.e., $d_1=3.7311$~{\AA}. The interlayer second-nearest-neighbor distance is the distance between atoms 9 and 3, i.e., $d_2=3.8520$~{\AA}. The interlayer third-nearest-neighbor distance is the distance between atoms 9 and 7, i.e., $d_3=4.9612$~{\AA}.

The original VFFM parameters from Ref~\onlinecite{KanetaC1986jpsj} are listed in the second column in Tab.~\ref{tab_vffm}. These parameters are further optimized in the present work, and are also listed in Tab.~\ref{tab_vffm}. After optimization, frequencies for the interlayer B mode and C mode in bulk BP agree well with the experimental results as shown in Tab.~\ref{tab_frequency}. 

The phonon modes are calculated using GULP.\cite{gulp} Tab.~\ref{tab_frequency} lists the frequency for the B mode and two C modes in bulk BP.  Due to the intrinsic geometric anisotropy due to the puckered configuration, the frequency for the interlayer shear mode in the x-direction (C$_{x}$) is quite different from the frequency of the interlayer shear mode in the y-direction (C$_{y}$). The frequencies of the interlayer modes from the optimized VFFM parameters agree quite well with the experiment, with a maximum error of about 3\% for the C$_{x}$ mode. The eigenvectors for the interlayer modes in the bulk BP are shown in Fig.~\ref{fig_bulk_all}. The figure is produced using XCRYSDEN.\cite{xcrysden}  

Tab.~\ref{tab_vffm} shows that $\alpha_2$ is negative in both the original VFFM parameter set and the optimized parameter set. A negative VFFM parameter implies that BP becomes unstable under high pressure. This pressure induced structure instability was investigated experimentally by Yamada et al.\cite{YamadaY1984prb}

\begin{figure}[tb]
  \begin{center}
    \scalebox{0.65}[0.65]{\includegraphics[width=8.0cm]{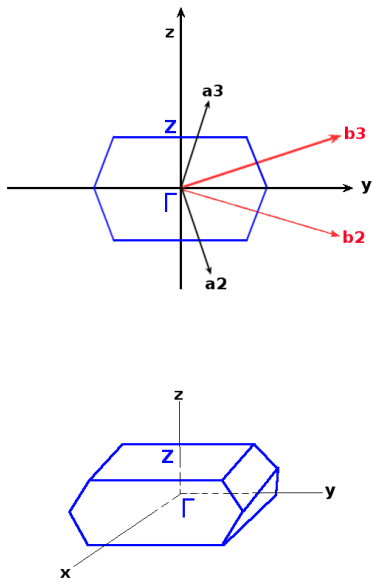}}
  \end{center}
  \caption{(Color online) The first Brillouin zone for bulk BP. Top is the projection of the first Brillouin zone onto the $yz$ plane. Bottom is the three-dimensional first Brillouin zone for bulk BP.}
  \label{fig_BZ}
\end{figure}

\section{Symmetry Analysis}
\subsection{Bulk BP}
There are eight atoms in the orthorhombic cell for bulk BP. The bases for the orthorhombic cell are as follows,
\begin{eqnarray}
\vec{A}_{1} & = & a\hat{e}_{x},\\
\vec{A}_{2} & = & b\hat{e}_{y},\\
\vec{A}_{3} & = & c\hat{e}_{z},
\end{eqnarray}
where $\hat{e}_{x}$, $\hat{e}_{y}$, and $\hat{e}_{z}$ are unit vectors in the three cartesian directions. The lattice constants $a=4.1766$~{\AA} and $b=3.2197$~{\AA} are computed from the Stillinger-Weber potential.\cite{JiangJW2014bpsw} The VFFM used for the interlayer interaction is linear, so it cannot be used to optimize the interlayer structure. We thus take the value of the lattice constant $c=10.587$~{\AA} from Ref~\onlinecite{DuY2010jap}.

\begin{table}
\caption{The irreducible representation for the $D_{2h}$ group at the Z point in the Brillouin zone. The two symbols for the irreducible representation are listed in the first column.}
\label{tab_d2h}
\begin{tabular}{@{\extracolsep{\fill}}|c|c|c|c|c|c|c|c|c|}
\hline 
 & E & $C_{2z}$ & $C_{2y}$ & $C_{2x}$ & $i$ & $\sigma_{xy}$ & $\sigma_{xz}$ & $\sigma_{yz}$\tabularnewline
\hline 
$A_{g}$=$Z_{2}^{+}$ & 1 & 1 & 1 & 1 & 1 & 1 & 1 & 1\tabularnewline
\hline 
$B_{1g}$=$Z_{3}^{+}$ & 1 & 1 & -1 & -1 & 1 & 1 & -1 & -1\tabularnewline
\hline 
$B_{2g}$=$Z_{1}^{+}$ & 1 & -1 & 1 & -1 & 1 & -1 & 1 & -1\tabularnewline
\hline 
$B_{3g}$=$Z_{4}^{+}$ & 1 & -1 & -1 & 1 & 1 & -1 & -1 & 1\tabularnewline
\hline 
$A_{u}$=$Z_{2}^{-}$ & 1 & 1 & 1 & 1 & -1 & -1 & -1 & -1\tabularnewline
\hline 
$B_{1u}$=$Z_{3}^{-}$ & 1 & 1 & -1 & -1 & -1 & -1 & 1 & 1\tabularnewline
\hline 
$B_{2u}$=$Z_{1}^{-}$ & 1 & -1 & 1 & -1 & -1 & 1 & -1 & 1\tabularnewline
\hline 
$B_{3u}$=$Z_{4}^{-}$ & 1 & -1 & -1 & 1 & -1 & 1 & 1 & -1\tabularnewline
\hline 
\end{tabular}
\end{table}

\begin{table*}
\caption{Symmetry analysis for phonon modes in bulk BP ($\Gamma$ point and Z point) and FLBP ($\Gamma$ point).  The total number of phonons is listed in the third column. Phonon modes are classified by the irreducible representations of $\Gamma^{vib}$ in the fourth column. The irreducible representations of the IR and Raman-active modes are listed in the fifth and sixth columns, respectively. $N$ is the layer number.}
\label{tab_allmode}
\begin{tabular*}{\textwidth}{@{\extracolsep{\fill}}|c|c|c|c|c|c|}
\hline 
 & group & mode number & $\Gamma^{vib}$ & $\Gamma^{IR}$ & $\Gamma^{R}$\tabularnewline
\hline 
\multirow{2}{*}{bulk at $\Gamma$ point} & \multirow{2}{*}{$D_{2h}$} & \multirow{2}{*}{12} & $2A_{g}\oplus B_{1g}\oplus2B_{2g}\oplus B_{3g}\oplus$ & \multirow{2}{*}{$B_{1u}\oplus B_{2u}\oplus B_{3u}$} & \multirow{2}{*}{$3A_{g}\oplus B_{1g}\oplus B_{2g}\oplus B_{3g}$}\tabularnewline
 &  &  & $A_{u}\oplus2B_{1u}\oplus B_{2u}\oplus2B_{3u}$ &  & \tabularnewline
\hline 
\multirow{2}{*}{bulk at Z point} & \multirow{2}{*}{$D_{2h}$} & \multirow{2}{*}{12} & $2Z_{2}^{+}\oplus Z_{3}^{+}\oplus2Z_{1}^{+}\oplus Z_{4}^{+}\oplus$ & \multirow{2}{*}{$Z_{1}^{-}\oplus Z_{3}^{-}\oplus Z_{4}^{-}$} & \multirow{2}{*}{$3Z_{2}^{+}\oplus Z_{3}^{+}\oplus Z_{1}^{+}\oplus Z_{4}^{+}$}\tabularnewline
 &  &  & $Z_{2}^{-}\oplus2Z_{3}^{-}\oplus Z_{1}^{-}\oplus2Z_{4}^{-}$ &  & \tabularnewline
\hline 
FLBP, even N & $C_{s}$ & $12N$ & $8NA'\oplus4NA''$ & $2A'\oplus A''$ & $4A'\oplus2A''$\tabularnewline
\hline 
FLBP, odd N & $C_{2h}$ & $12N$ & $4NA_{g}\oplus2NB_{g}\oplus2NA_{u}\oplus4NB_{u}$ & $A_{u}\oplus2B_{u}$ & $4A_{g}\oplus2B_{g}$\tabularnewline
\hline 
\end{tabular*}
\end{table*}

\begin{table*}
\caption{The irreducible representation for the B mode and C modes. IR or Raman-activity is listed in the parentheses, where `No' indicates optically inactive. $N$ is the layer number.}
\label{tab_bcmode}
\begin{tabular*}{\textwidth}{@{\extracolsep{\fill}}|c|c|c|c|c|c|c|}
\hline 
 & $B_{1}$ mode & $C_{x1}$ mode & $C_{y1}$ mode & $B_{2}$ mode & $C_{x2}$ mode & $C_{y2}$ mode\tabularnewline
\hline 
bulk & $Z_{4}^{-}$ (IR) & $Z_{3}^{-}$ (IR) & $Z_{2}^{-}$ (No) & / & / & /\tabularnewline
\hline 
FLBP, even N & $A'$ (IR, R) & $A'$ (IR, R) & $A''$ (IR, R) & $A'$ (IR, R) & $A'$ (IR, R) & $A''$ (IR, R)\tabularnewline
\hline 
FLBP, odd N & $A_{g}$ (R) & $B_{u}$ (IR) & $A_{u}$ (IR) & $B_{u}$ (IR) & $A_{g}$ (R) & $B_{g}$ (R)\tabularnewline
\hline 
Layer dependence & $\sqrt{1-\cos\frac{\pi}{N}}$ & $\sqrt{1+\cos\frac{\pi}{N}}$ & $\sqrt{1+\cos\frac{\pi}{N}}$ & $\sqrt{1-\cos\frac{2\pi}{N}}$ & $\sqrt{1+\cos\frac{2\pi}{N}}$ & $\sqrt{1+\cos\frac{2\pi}{N}}$\tabularnewline
\hline 
\end{tabular*}
\end{table*}

The primitive unit cell for the bulk BP contains four atoms.\cite{SlaterJC1962pr} The bases for the primitive unit cell are,
\begin{eqnarray}
\vec{a}_{1} & = & \vec{A}_{1},\\
\vec{a}_{2} & = & \frac{1}{2}\left(\vec{A}_{2}-\vec{A}_{3}\right),\\
\vec{a}_{3} & = & \frac{1}{2}\left(\vec{A}_{2}+\vec{A}_{3}\right).
\end{eqnarray}
Each unit cell can be labelled by a lattice vector $\vec{R}_{l_1l_2l_3}=l_1\vec{a}_{1} + l_2\vec{a}_2 + l_3\vec{a}_3$, with $l_1$, $l_2$, and $l_3$ as three integers. The lattice vector corresponds to a translation symmetry operation, $\hat{T}_{l_1l_2l_3}$, which translates the bulk BP by a lattice vector $\vec{R}_{l_1l_2l_3}$.

The point group for the bulk BP is D$_{2h}$ = \{$E$, C$_{2z}$, C$_{2y}$,C$_{2x}$, $i$, $\sigma_{xy}$, $\sigma_{yz}$, $\sigma_{zx}$\}. C$_{2z}$ is the rotation for $\pi$ around the z-axis, $i$ is the inversion symmetry and $\sigma_{xy}$ is the reflection with respect to the $z=0$ plane. Four of these eight symmetry operations, C$_{2z}$, C$_{2x}$, $\sigma_{xy}$, and  $\sigma_{yz}$, are accompanied by the following nonprimitive translations,
\begin{eqnarray}
\vec{\tau} & = & \frac{1}{2}\left(\vec{A}_{1}+\vec{A}_{3}\right).
\end{eqnarray}
The translational symmetry and the point group together construct the space group (D$_{2h}^{18}$) of bulk BP; i.e., D$_{2h}^{18}$ = $\hat{T}_{l_1l_2l_3}$ $\otimes$ D$_{2h}$.

The reciprocal vectors are determined by the bases for the primitive unit cell through the relation
\begin{eqnarray}
\vec{b}_{i}\cdot \vec{a}_{j} = 2\pi \delta_{ij},
\end{eqnarray}
which gives,
\begin{eqnarray}
\vec{b}_{1} & = & \frac{2\pi}{a}\hat{e}_{x},\\
\vec{b}_{2} & = & \frac{2\pi}{b}\hat{e}_{y}-\frac{2\pi}{c}\hat{e}_{z},\\
\vec{b}_{3} & = & \frac{2\pi}{b}\hat{e}_{y}+\frac{2\pi}{c}\hat{e}_{z}.
\end{eqnarray}
$\vec{b}_{1}$ is in x-direction, while $\vec{b}_{2}$ and $\vec{b}_{3}$ lie in the yz plane. The first Brillouin zone for bulk BP is shown in Fig.~\ref{fig_BZ}.

The Z point in the first Brillouin zone plays an important role in the present work. The wave vector for the B mode and C mode in the bulk BP is located at the Z point of the first Brillouin zone, and not at the $\Gamma$ point of the first Brillouin zone. This can be demonstrated as follows. The wave vector for the Z point is,
\begin{eqnarray}
\vec{k}_{Z} & = & \frac{1}{2}\left(-\vec{b}_{2}+\vec{b}_{3}\right).
\end{eqnarray}
We shall treat the unit cell containing the four atoms (1, 2, 3, 4) in Fig.~\ref{fig_cfg} as the (0, 0, 0) unit cell. The lattice vector for this unit cell is $\vec{R}_{000}=0$. This unit cell (0, 0, 0) is in the same plane as the unit cell containing atoms (5, 6, 7, 8). The lattice vector for the latter unit cell is $\vec{R}_{011}=\vec{A}_{2}=\vec{a}_{2}+\vec{a}_{3}$, so its phase factor in the Bloch theory is $\vec{k}_{Z}\cdot \vec{R}_{011}=0$, which means that the phase factors for the unit cells in the same BP plane are the same. The lattice vector for the unit cell containing atoms (9, 10, 11, 12) is $\vec{R}_{010}=\vec{a}_{2}$, with the phase factor as $\vec{k}_{Z}\cdot \vec{R}_{010}=-\pi$. This unit cell (0, 1, 0) is in a different layer from the (0, 0, 0) unit cell. It shows that the vibration for the two BP layers in bulk BP are out-of-phase at the Z point. We therefore have demonstrated that the phonon modes at the Z point correspond to the relative vibrations between the two BP layers in bulk BP. The B mode and C mode studied in the present work describe the relative breathing or shearing motion of the two BP layers, so the wave vectors for these modes are located at the Z point.

There are twelve phonon modes at the $\Gamma$ point or Z point for bulk BP, corresponding to the four atoms in the primitive unit cell. The symmetry for these phonon modes can be analyzed according to the point group (D$_{2h}$) of bulk BP. Tab.~\ref{tab_d2h} lists the eight irreducible representations for the point group D$_{2h}$. There are two symbols for each irreducible representation in the first column of Tab.~\ref{tab_d2h}. The first symbol is for phonon modes at the $\Gamma$ point, while the second symbol is for the phonon modes at the Z point in the first Brillouin zone.

Tab.~\ref{tab_allmode} shows the symmetry analysis for phonon modes at the $\Gamma$ point or the Z point in the first Brillouin zone. $\Gamma^{vib}=\Gamma^{a.s}\otimes\Gamma^{vec}$ is the vibrational representation, with $\Gamma^{a.s}$ as the permutation representation and $\Gamma^{vec}$ as the vector representation. The decomposition of the vibrational representation gives the irreducible representation for each phonon mode at the Z point in the first Brillouin zone. The vibrational representation can be decomposed as follows, using the character table method,
\begin{eqnarray}
\Gamma^{vib}  &=& 2Z_2^+\oplus Z_3^+\oplus2Z_1^+\oplus Z_4^+\nonumber\\
&&\oplus Z_2^-\oplus2Z_3^-\oplus Z_1^-\oplus2Z_4^-.
\label{eq_vib_bulk}
\end{eqnarray}
The twelve phonon modes at the Z point in the first Brillouin zone belong to these 12 irreducible representations on the right-hand side of Eq.~(\ref{eq_vib_bulk}). From the eigenvector, it can be determined that the B mode in bulk BP belongs to the $Z_4^-$ irreducible representation, the C$_{x}$ mode belongs to the $Z_3^-$ irreducible representation, and the C$_{y}$ mode belongs to the $Z_2^-$ irreducible representation. These results are shown in the second line of Tab.~\ref{tab_bcmode}.

The IR activity for each phonon mode can be analyzed by decomposing the vector representation in the following,
\begin{eqnarray}
\Gamma^{vec}  = Z_{1}^{-}\oplus Z_{3}^{-}\oplus Z_{4}^{-}.
\end{eqnarray}
The vector representation is three-dimensional, so there are three one-dimensional irreducible representations in the resulting decomposition. This result predicts that phonon modes corresponding to these three irreducible representations will be IR-active in the optical scattering process. According to this result, the B mode ($Z_4^-$) and C$_x$ mode ($Z_3^-$) in bulk BP are IR-active, while the C$_y$ mode ($Z_2^-$) is IR inactive.

The Raman activity for each phonon mode can be determined by decomposing the six-dimensional tensor representation, $\Gamma^{v\times v}$. The bases for the tensor representation are $x^{2}+y^{2}$ , $z^{2}$, $x^{2}-y^{2}$, $xy$ , $xz$ , and $yz$. The tensor representation is decomposed as follows,
\begin{eqnarray*}
\Gamma^{v\times v} = 3Z_{2}^{+}\oplus Z_{3}^{+}\oplus Z_{1}^{+}\oplus Z_{4}^{+}.
\end{eqnarray*}
According to this decomposition result, none of the three interlayer modes is Raman active in bulk BP.

The above decomposition results for phonon modes at the Z point in bulk BP are shown in the third line of Tab.~\ref{tab_allmode}. The symmetry analysis for phonon modes at the $\Gamma$ point in the first Brillouin zone of bulk BP are shown in the second line of Tab.~\ref{tab_allmode}. The symmetry for the three interlayer modes for bulk BP are shown in the second line of Tab.~\ref{tab_bcmode}.

\subsection{Few-Layer BP}

\begin{table}
\caption{The irreducible representation for the $C_{s}$ group.}
\label{tab_cs}
\begin{tabular}{@{\extracolsep{\fill}}|c|c|c|}
\hline 
 & E & $\sigma_{h}$\tabularnewline
\hline 
$A'$ & 1 & 1\tabularnewline
\hline 
$A''$ & 1 & -1\tabularnewline
\hline 
\end{tabular}
\end{table}

\begin{table}
\caption{The irreducible representation for the $C_{2h}$ group.}
\label{tab_c2h}
\begin{tabular}{@{\extracolsep{\fill}}|c|c|c|c|c|}
\hline 
 & E & $C_{2}$ & i & $\sigma_{h}$\tabularnewline
\hline 
$A_{g}$ & 1 & 1 & 1 & 1\tabularnewline
\hline 
$B_{g}$ & 1 & -1 & 1 & -1\tabularnewline
\hline 
$A_{u}$ & 1 & 1 & -1 & -1\tabularnewline
\hline 
$B_{u}$ & 1 & -1 & -1 & 1\tabularnewline
\hline 
\end{tabular}
\end{table}

There is no translational symmetry in the z-direction for FLBP. Hence, the orthorhombic cell for the FLBP is the primitive unit cell in this structure. The bases are $\vec{A}_2$ and $\vec{A}_3$. There are $4N$ atoms in the primitive unit cell of the FLBP, whose symmetry is greatly reduced as compared with bulk BP.  Specifically, the four symmetry operations with nonprimitive translations in bulk BP are forbidden in FLBP.

For even-layer FLBP, only the $\sigma_{zx}$ plane reflection is a symmetry operation. The point group for the FLBP with even $N$ is $C_s$. The character table for the $C_s$ group is shown in Tab.~\ref{tab_cs}. The irreducible representation for each phonon mode at the $\Gamma$ point is found by decomposing the vibrational representation for the FLBP with even $N$. The symmetry analysis results are shown in the fourth line of Tab.~\ref{tab_allmode}. All phonon modes at the $\Gamma$ point are both IR-active and Raman-active, due to the low symmetry of this structure. In particular, there is no inversion symmetry in FLBP with even $N$, so the phonon mode can be IR-active and Raman-active simultaneously.

For the B mode, we are interested in the first lowest-frequency B (B$_1$) mode and the second lowest-frequency B (B$_2$) mode in FLBP. The eigenvectors for these two B modes are shown in Fig.~\ref{fig_u_b1_b2}. For the C mode, we are interested in the first highest-frequency C mode (C$_{x1}$ or C$_{y1}$) and the second highest-frequency C (C$_{x2}$ or C$_{y2}$) mode. The eigenvectors for these C modes are shown in Figs.~\ref{fig_u_cx1_cx2} and \ref{fig_u_cy1_cy2}. The third line of Tab.~\ref{tab_bcmode} shows the symmetry for these modes in the FLBP with even $N$.

For odd-layer FLBP, the point group is $C_{2h}$ = \{$E$, $C_{2y}=C_{2}$, $i$, $\sigma_{zx}=\sigma_{h}$\}. Tab.~\ref{tab_c2h} lists the character for the point group $C_{2h}$. The symmetry analysis for phonon modes at $\Gamma$ point in the odd-layer FLBP are shown in the fifth line of Tab.~\ref{tab_allmode}. The symmetry for the interlayer modes are shown in the fourth line of Tab.~\ref{tab_bcmode}. The IR-activity and Raman-activity are also shown in the table.

\section{Numerical Results}
For the interlayer mode in the layered structure, each layer can be regarded as a single atom. The whole layered structure can thus be considered as a single atomic chain with free boundary conditions at the two ends. This atomic chain model has been successfully applied to simulate the interlayer modes in few-layer graphene\cite{TanPH} and few-layer MoS$_2$.\cite{ZhaoYY} It can be assumed that each atom in the chain only interacts with its nearest-neighboring atoms. The eigenvector for the phonon mode $\tau$ in the chain model is
\begin{eqnarray}
u_{j}^{\tau} & \propto & \cos\left[\frac{\left(\tau-1\right)\left(2j-1\right)\pi}{2N}\right],
\label{eq_chain_u}
\end{eqnarray}
where $\tau$ is the mode index, $N$ is the total atom number and $j$ is the site index for each atom. The first mode ($\tau=1$) is the acoustic mode. The frequency for mode $\tau$ is
\begin{eqnarray}
\omega_{\tau} & = & \sqrt{\frac{\beta}{2\mu\pi^{2}c^{2}}\left\{ 1-\cos\left[\frac{\left(\tau-1\right)\pi}{N}\right]\right\} }.
\label{eq_chain_omega}
\end{eqnarray}
$\mu=1.53\times 10^{-26}$~{kg\AA$^{-2}$} is the mass per unit area of the single-layer BP, $c$ is the speed of light in {cm/s} and $\beta$ is the force constant per unit area.

\begin{figure*}[tb]
  \begin{center}
    \scalebox{0.85}[0.85]{\includegraphics[width=\textwidth]{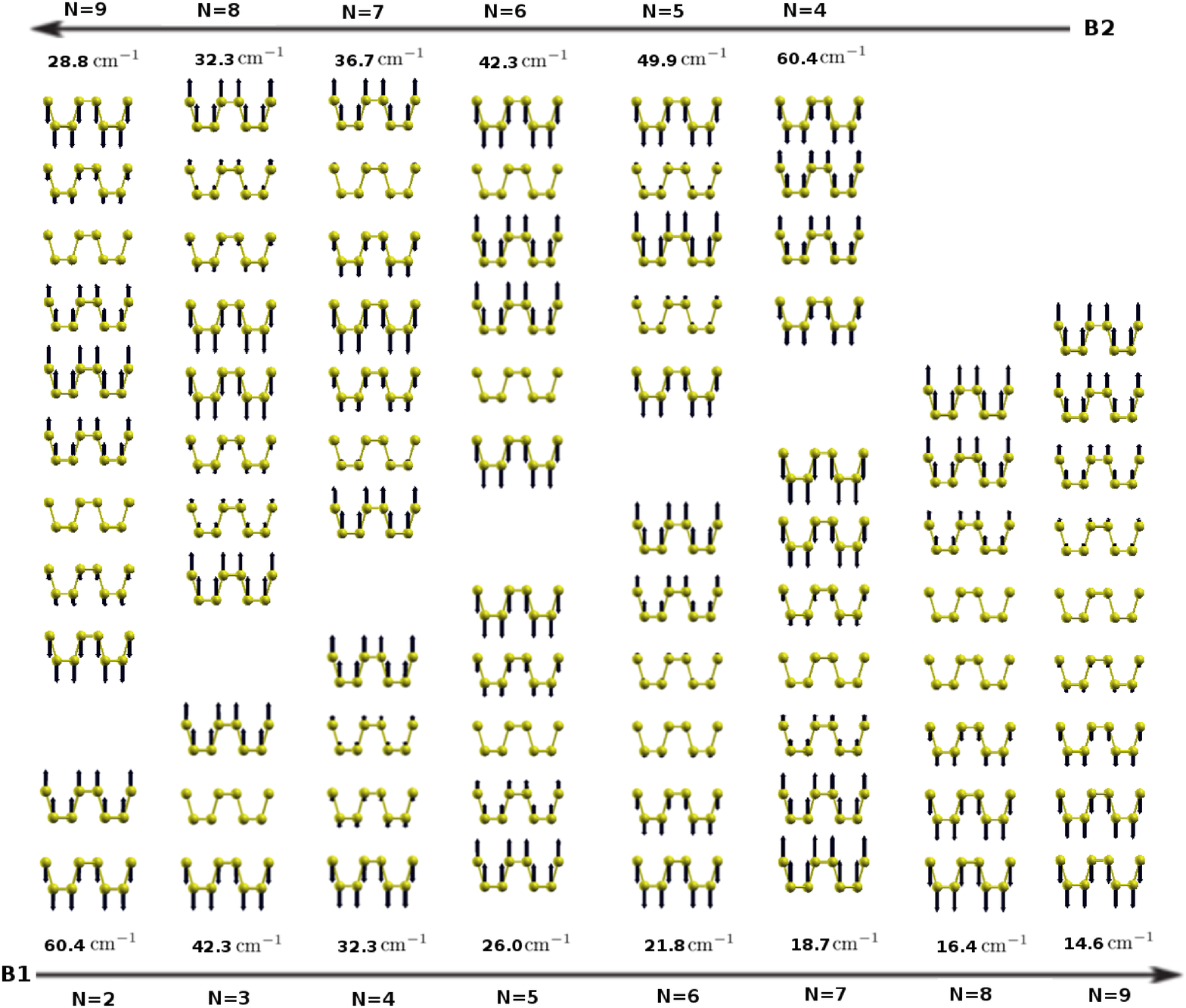}}
  \end{center}
  \caption{(Color online) Eigenvectors and frequencies for two B modes. Bottom is the lowest-frequency B mode. Top is the second-lowest-frequency B mode.}
  \label{fig_u_b1_b2}
\end{figure*}

\begin{figure*}[tb]
  \begin{center}
    \scalebox{0.85}[0.85]{\includegraphics[width=\textwidth]{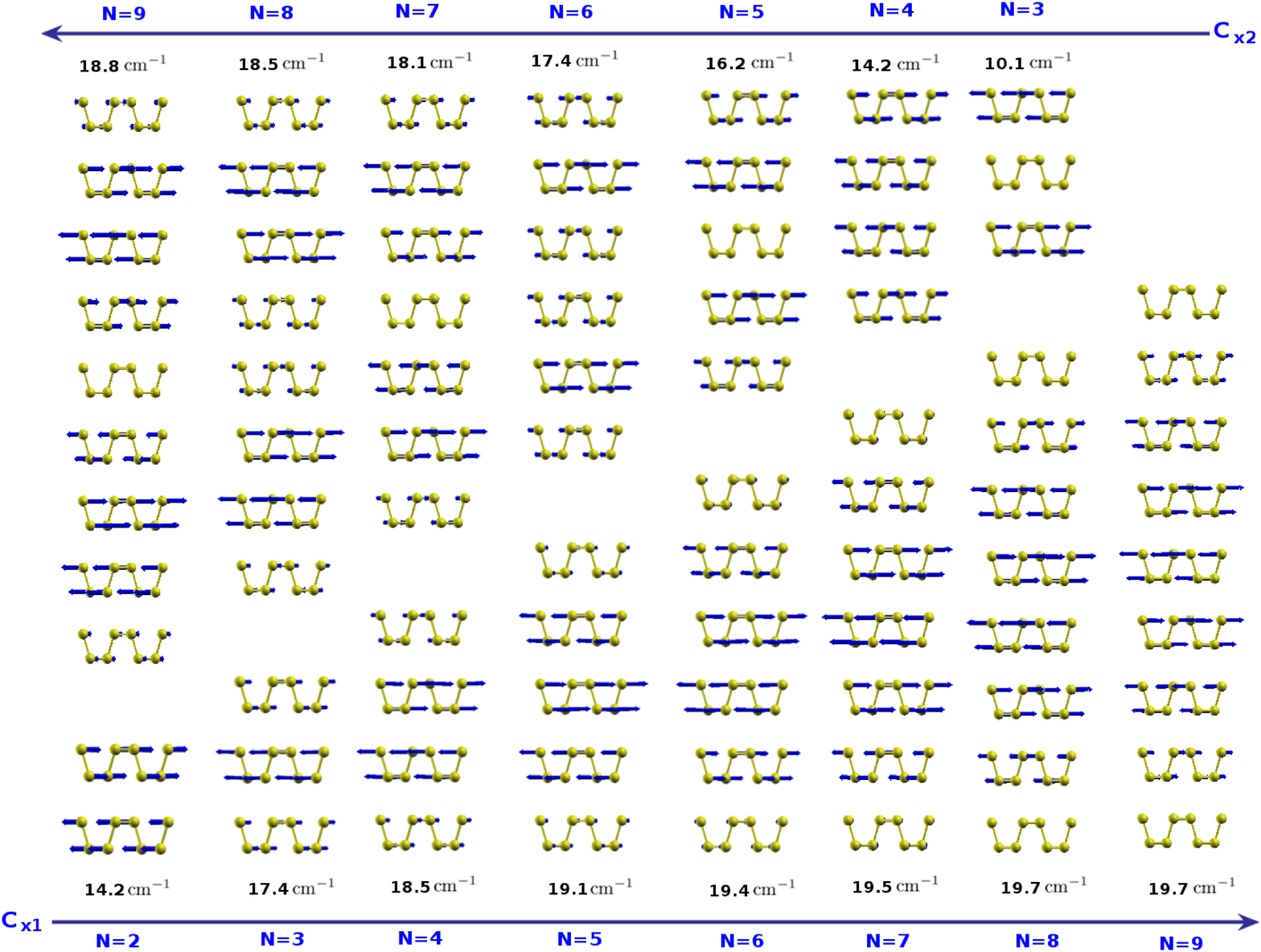}}
  \end{center}
  \caption{(Color online) Eigenvectors and frequencies for two C$_x$ modes. Bottom is the highest-frequency C$_x$ mode. Top is the second-highest-frequency C$_x$ mode.}
  \label{fig_u_cx1_cx2}
\end{figure*}

\begin{figure*}[tb]
  \begin{center}
    \scalebox{0.85}[0.85]{\includegraphics[width=\textwidth]{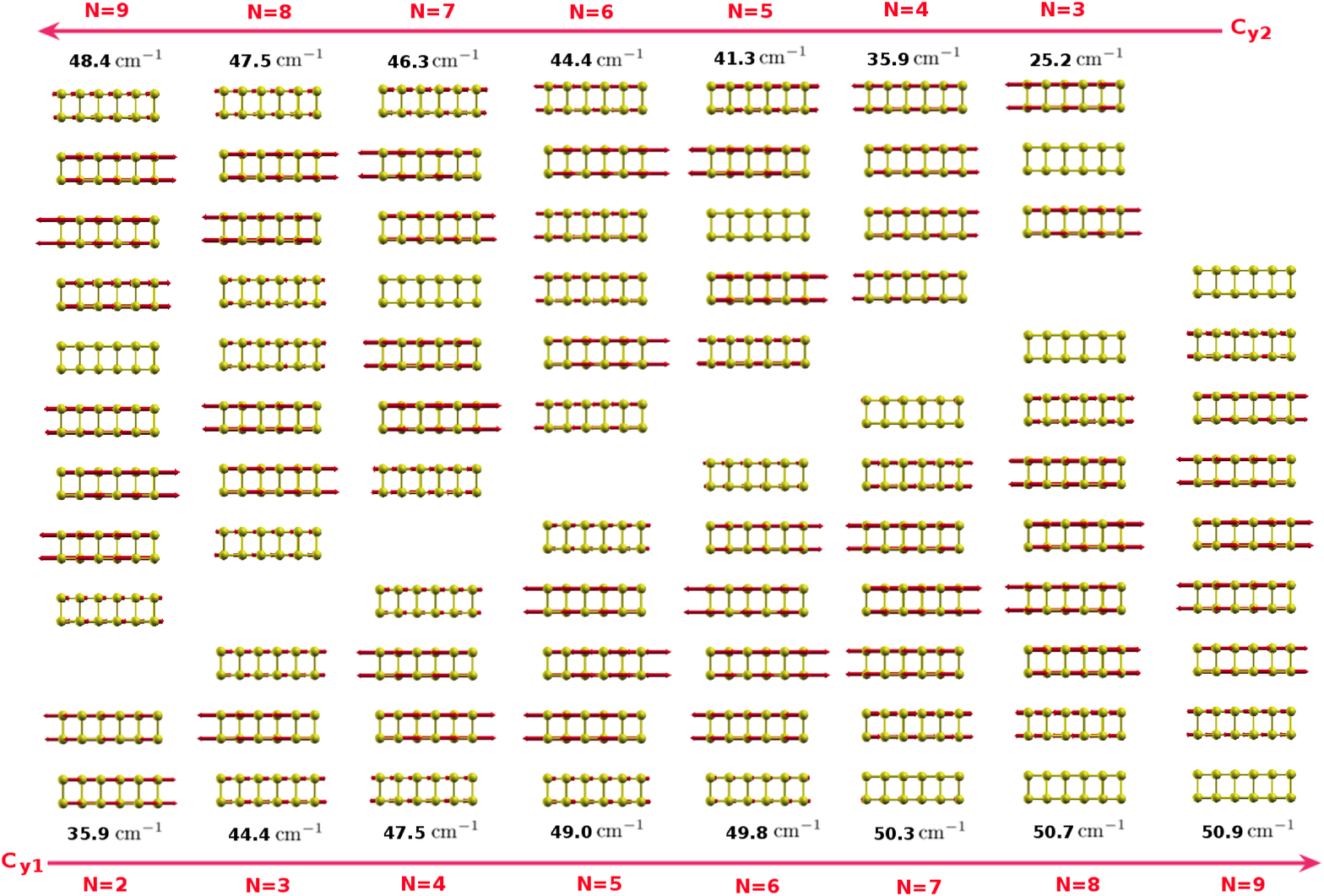}}
  \end{center}
  \caption{(Color online) Eigenvectors and frequencies for two C$_y$ modes. Bottom is the highest-frequency C$_{y1}$ mode. Top is the second-highest-frequency C$_{y2}$ mode.}
  \label{fig_u_cy1_cy2}
\end{figure*}

\begin{figure}[tb]
  \begin{center}
    \scalebox{0.85}[0.85]{\includegraphics[width=8cm]{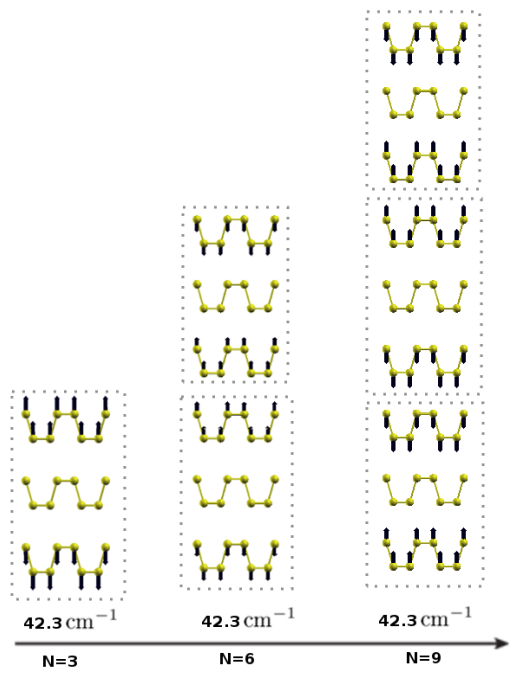}}
  \end{center}
  \caption{(Color online) Interlayer collective B modes with the same frequency for FLBP with $N=$ 3, 6, and 9 (from left to right). The eigenvector can be regarded as the collective vibration of the small segments (dotted rectangles).}
  \label{fig_b_same_frequency}
\end{figure}

\begin{figure}[tb]
  \begin{center}
    \scalebox{0.85}[0.85]{\includegraphics[width=8cm]{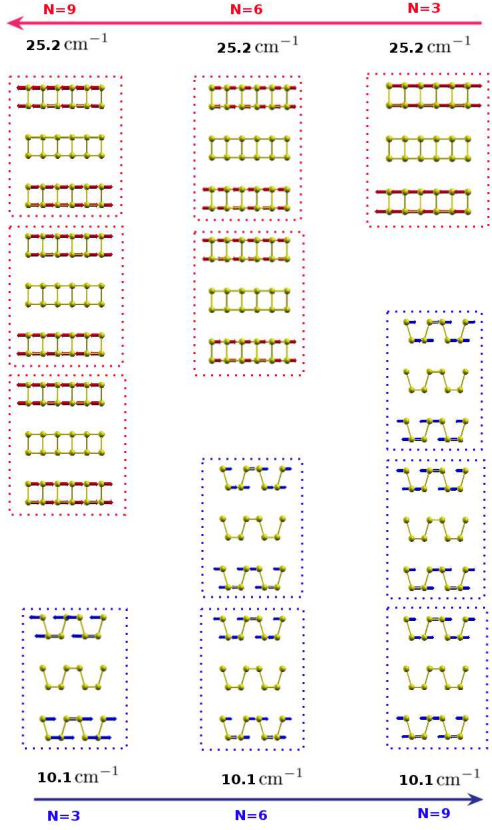}}
  \end{center}
  \caption{(Color online) Interlayer collective C modes with the same frequency for FLBP with $N=$ 3, 6, and 9. Bottom is the C$_x$ mode, and  top is the C$_y$ mode.}
  \label{fig_c_same_frequency}
\end{figure}

\begin{figure}[tb]
  \begin{center}
    \scalebox{1}[1]{\includegraphics[width=8.0cm]{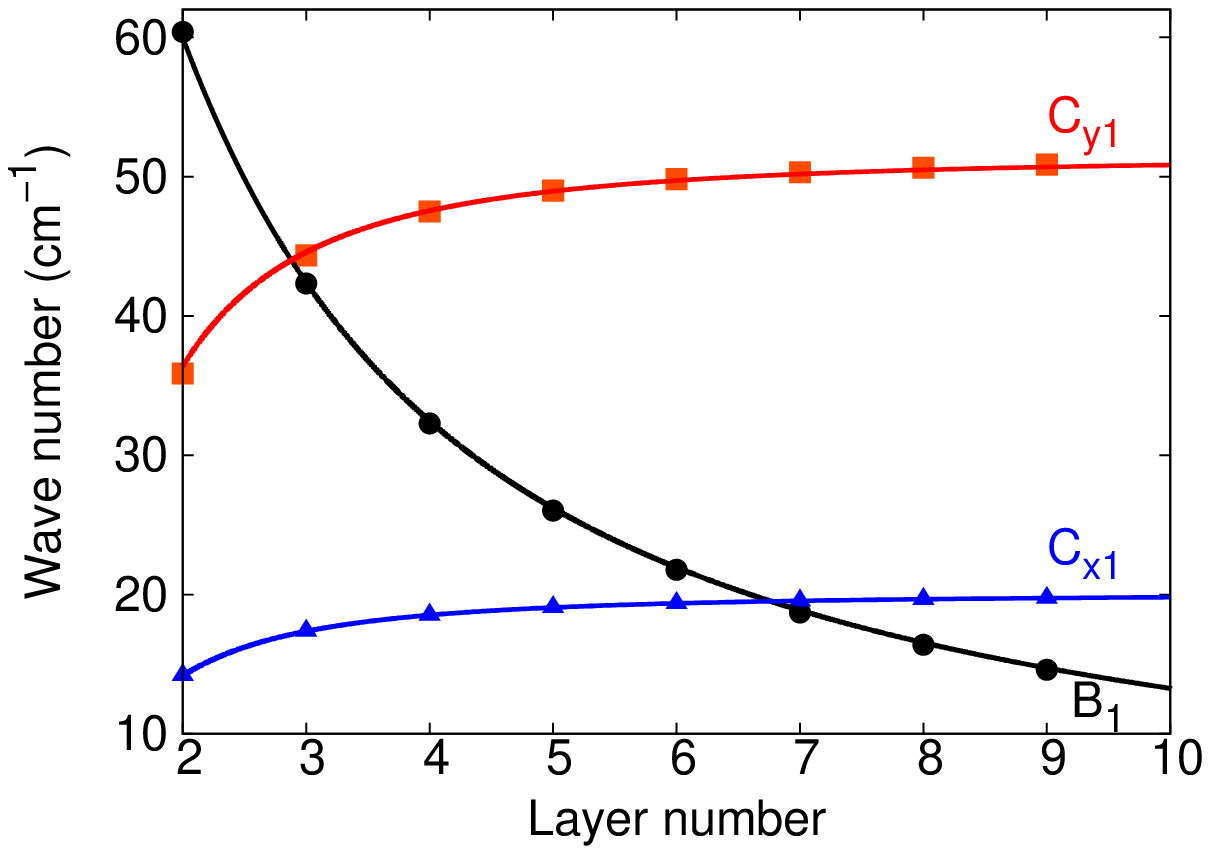}}
  \end{center}
  \caption{(Color online) The layer dependence for some interlayer modes. B$_1$ is the lowest-frequency interlayer B mode. C$_{x1}$ is the highest-frequency interlayer C mode in the x-direction. C$_{y1}$ is the highest-frequency interlayer C mode in the y-direction. Lines are fitting functions according to the chain model.}
  \label{fig_frequency1}
\end{figure}

\begin{figure}[tb]
  \begin{center}
    \scalebox{1}[1]{\includegraphics[width=8.0cm]{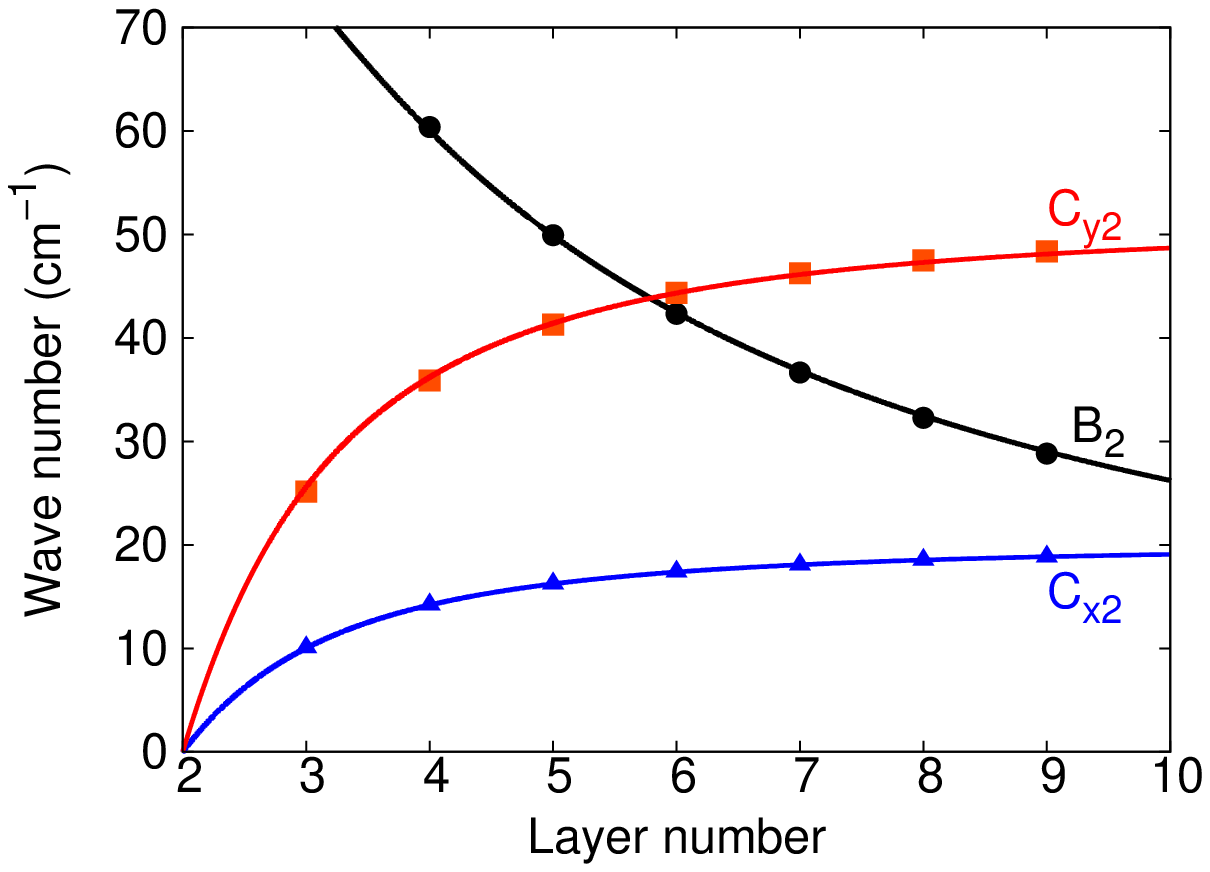}}
  \end{center}
  \caption{(Color online) The layer dependence for some interlayer modes. B$_2$ is the second lowest-frequency interlayer B mode. C$_{x2}$ is the second highest-frequency interlayer C mode in the x-direction. C$_{y2}$ is the second highest-frequency interlayer C mode in the y-direction. Lines are fitting functions according to the chain model.}
  \label{fig_frequency2}
\end{figure}

We discuss four sets of interlayer phonon modes for FLBP in this section. The eigenvectors for these modes are shown in Figs.~\ref{fig_u_b1_b2}~-~\ref{fig_c_same_frequency}. The layer number dependence for the frequency are shown in Figs.~\ref{fig_frequency1} and \ref{fig_frequency2}.

The first set is the two B modes, i.e., B$_1$ mode and B$_2$ mode. The B$_1$ mode corresponds to the phonon mode with $\tau=2$ in the chain model. Fig.~\ref{fig_u_b1_b2} shows that the eigenvector of the B$_1$ mode indeed follows the prediction of the chain model, i.e., $u_{j}^{2} \propto \cos\left[\frac{\left(2j-1\right)\pi}{2N}\right]$. The $N$-dependence for the frequency of the B$_1$ mode is shown in Fig.~\ref{fig_frequency1}, where the black solid line illustrates a perfect fitting of the frequency for the B$_1$ mode to the function $\omega_{2} = \sqrt{\frac{\beta_{B}}{2\mu\pi^{2}c^{2}}\left[ 1-\cos\left(\frac{\pi}{N}\right)\right] }$. These results show that the B$_1$ mode can be well described by the chain model. From the fitting of the frequency, we get the breathing force constant parameter $\beta_B=9.8\times 10^{19}$~{Nm$^{-3}$} for the chain model.

The B$_2$ mode corresponds to $\tau=3$ in the chain model. Its eigenvector is shown in Fig.~\ref{fig_u_b1_b2}, which agrees with the prediction of the chain model, i.e., $u_{j}^{3} \propto \cos\left[\frac{\left(2j-1\right)\pi}{N}\right]$. The $N$-dependence for the frequency of B$_2$ mode is shown in Fig.~\ref{fig_frequency2}, where the black solid line shows that the frequency for the B$_2$ mode can be well fitted to the function $\omega_{3} = \sqrt{\frac{\beta_{B}}{2\mu\pi^{2}c^{2}}\left[ 1-\cos\left(\frac{2\pi}{N}\right)\right] }$. The fitting parameter $\beta_B=9.8\times 10^{19}$~{Nm$^{-3}$} is exactly the same as that obtained from the B$_1$ mode. This agreement further confirms the success of the chain model in the description of the layered structure.

The second set of phonon modes are the interlayer C$_x$ modes in FLBP, including the first highest-frequency C$_{x1}$ mode and the second highest-frequency C$_{x2}$ mode. This mode can also be described by the chain model. The C$_{x1}$ mode corresponds to the phonon mode with $\tau=N$ in the chain model. The eigenvector of the C$_{x1}$ mode is shown in Fig.~\ref{fig_u_cx1_cx2}. This eigenvector follows the function $u_{j}^{N} \propto \cos\left[\frac{\left(N-1\right)\left(2j-1\right)\pi}{2N}\right]$. The $N$-dependence for the frequency of the C$_{x1}$ mode is shown in Fig.~\ref{fig_frequency1}, which is fitted to the function $\omega_{N} = \sqrt{\frac{\beta_{Cx}}{2\mu\pi^{2}c^{2}}\left[ 1+\cos\left(\frac{\pi}{N}\right)\right] }$. The fitting parameter $\beta_{Cx}=5.5\times 10^{18}$~{Nm$^{-3}$} is the force constant for the transverse motion in the x-direction for the chain model.

The C$_{x2}$ mode corresponds to $\tau=N-1$ mode in the chain model. Its eigenvector is shown in Fig.~\ref{fig_u_cx1_cx2}, which coincides with the eigenvector of the chain model with $\tau=N-1$ in Eq.~(\ref{eq_chain_u}). The frequency for the C$_{x2}$ is shown in Fig.~\ref{fig_frequency2}. The $N$-dependence of the frequency is also consistent with the chain model prediction by Eq.~(\ref{eq_chain_omega}) with $\tau=N-1$.

The third set of phonon modes are two interlayer C$_y$ modes in FLBP; i.e., the first highest frequency C$_{y1}$ mode and the second highest-frequency C$_{y2}$ mode. These modes can also be described by the chain model. C$_{y1}$ mode corresponds to the phonon mode with $\tau=N$ in the chain model, while the C$_{y2}$ mode corresponds to the phonon mode with $\tau=N-1$ in the chain model. The eigenvectors of these two modes are shown in Fig.~\ref{fig_u_cy1_cy2}. They agree with the chain model prediction in Eq.~(\ref{eq_chain_u}). The frequencies of these two modes are shown in Fig.~\ref{fig_frequency1} and Fig.~\ref{fig_frequency2}. They can be fitted by the frequency in the chain model in Eq.~(\ref{eq_chain_omega}). The fitted parameter $\beta_{Cy}=3.6\times 10^{19}$~{Nm$^{-3}$} is the force constant for the y-directional transverse motion for the chain model. The transverse force constant is about $1.28\times 10^{19}$~{Nm$^{-3}$} for the few-layer graphene\cite{TanPH} and $2.7\times 10^{19}$~{Nm$^{-3}$} in the few-layer MoS$_{2}$.\cite{ZhaoYY} These values are sandwiched between the two transverse force constants $\beta_{Cx}$ and $\beta_{Cy}$ in FLBP.

The fourth set of phonon modes are the collective vibration modes. Fig.~\ref{fig_b_same_frequency} shows a particular B mode in FLBP with layer number $N=3i$ ($i=1, 2, 3, ..., $). The frequencies of these collective B modes are independent of the layer number. For $N=6$, the third and fourth BP layers have the same vibrational displacement, so the overall displacement can be regarded as a collective vibration of two segments (displayed by dotted rectangles). Furthermore, the displacement for each segment is the same as the displacement for $N=3$. As a result, the frequency for the phonon mode in 6-layer BP is the same as the frequency for 3-layer BP. Similarly, for $N=9$, the structure can be deconstructed into three collective segments. Each segment has the same displacement as FLBP with $N=3$, so the frequency for the phonon mode in FLBP with $N=9$ is the same as the FLBP with $N=3$. Fig.~\ref{fig_c_same_frequency} shows a similar phenomenon for the collective C$_x$ mode and C$_y$ modes in FLBP. As shown in Tab.~\ref{tab_bcmode}, this set of phonon modes are optically active, so we expect that it will be possible to identify these FLBP phonon modes experimentally.

\section{conclusion}

To summarize, we have analyzed the lattice dynamical properties for the interlayer B and C modes in FLBP. The symmetry group for the FLBP with even layer numbers is compared with the FLBP with odd layer numbers. The symmetry properties for the interlayer modes are determined based on the symmetry groups. The IR and Raman activity for the phonon modes is also determined. We applied the VFFM to compute the eigenvectors and frequencies for the interlayer modes, which can be successfully explained by the chain model. The two C modes have very different frequencies, due to the anisotropic puckered configuration for the BP layer. We found a particular set of collective phonon modes with the same frequency in the FLBP with layer number $N=3i$ ($i$ integer). These collective phonon modes have a constant frequency with respect to the layer number.

\textbf{Acknowledgements} The work is supported by the Recruitment Program of Global Youth Experts of China and the start-up funding from Shanghai University. HSP acknowledges the support of the Mechanical Engineering department at Boston University.


\begin{thebibliography}{21}%
\makeatletter
\providecommand \@ifxundefined [1]{%
 \@ifx{#1\undefined}
}%
\providecommand \@ifnum [1]{%
 \ifnum #1\expandafter \@firstoftwo
 \else \expandafter \@secondoftwo
 \fi
}%
\providecommand \@ifx [1]{%
 \ifx #1\expandafter \@firstoftwo
 \else \expandafter \@secondoftwo
 \fi
}%
\providecommand \natexlab [1]{#1}%
\providecommand \enquote  [1]{``#1''}%
\providecommand \bibnamefont  [1]{#1}%
\providecommand \bibfnamefont [1]{#1}%
\providecommand \citenamefont [1]{#1}%
\providecommand \href@noop [0]{\@secondoftwo}%
\providecommand \href [0]{\begingroup \@sanitize@url \@href}%
\providecommand \@href[1]{\@@startlink{#1}\@@href}%
\providecommand \@@href[1]{\endgroup#1\@@endlink}%
\providecommand \@sanitize@url [0]{\catcode `\\12\catcode `\$12\catcode
  `\&12\catcode `\#12\catcode `\^12\catcode `\_12\catcode `\%12\relax}%
\providecommand \@@startlink[1]{}%
\providecommand \@@endlink[0]{}%
\providecommand \url  [0]{\begingroup\@sanitize@url \@url }%
\providecommand \@url [1]{\endgroup\@href {#1}{\urlprefix }}%
\providecommand \urlprefix  [0]{URL }%
\providecommand \Eprint [0]{\href }%
\providecommand \doibase [0]{http://dx.doi.org/}%
\providecommand \selectlanguage [0]{\@gobble}%
\providecommand \bibinfo  [0]{\@secondoftwo}%
\providecommand \bibfield  [0]{\@secondoftwo}%
\providecommand \translation [1]{[#1]}%
\providecommand \BibitemOpen [0]{}%
\providecommand \bibitemStop [0]{}%
\providecommand \bibitemNoStop [0]{.\EOS\space}%
\providecommand \EOS [0]{\spacefactor3000\relax}%
\providecommand \BibitemShut  [1]{\csname bibitem#1\endcsname}%
\let\auto@bib@innerbib\@empty
\bibitem [{\citenamefont {Neto}\ and\ \citenamefont
  {Novoselov}(2011)}]{NetoAHC2011rpp}%
  \BibitemOpen
  \bibfield  {author} {\bibinfo {author} {\bibfnamefont {A.~H.~C.}\
  \bibnamefont {Neto}}\ and\ \bibinfo {author} {\bibfnamefont {K.}~\bibnamefont
  {Novoselov}},\ }\href@noop {} {\bibfield  {journal} {\bibinfo  {journal}
  {Rep. Prog. Phys.}\ }\textbf {\bibinfo {volume} {74}},\ \bibinfo {pages}
  {082501} (\bibinfo {year} {2011})}\BibitemShut {NoStop}%
\bibitem [{\citenamefont {Tan}\ \emph {et~al.}(2012)\citenamefont {Tan},
  \citenamefont {Han}, \citenamefont {Zhao}, \citenamefont {Wu}, \citenamefont
  {Chang}, \citenamefont {Wang}, \citenamefont {Wang}, \citenamefont {Bonini},
  \citenamefont {Marzari}, \citenamefont {Pugno}, \citenamefont {Savini},
  \citenamefont {Lombardo},\ and\ \citenamefont {Ferrari}}]{TanPH}%
  \BibitemOpen
  \bibfield  {author} {\bibinfo {author} {\bibfnamefont {P.~H.}\ \bibnamefont
  {Tan}}, \bibinfo {author} {\bibfnamefont {W.~P.}\ \bibnamefont {Han}},
  \bibinfo {author} {\bibfnamefont {W.~J.}\ \bibnamefont {Zhao}}, \bibinfo
  {author} {\bibfnamefont {Z.~H.}\ \bibnamefont {Wu}}, \bibinfo {author}
  {\bibfnamefont {K.}~\bibnamefont {Chang}}, \bibinfo {author} {\bibfnamefont
  {H.}~\bibnamefont {Wang}}, \bibinfo {author} {\bibfnamefont {Y.~F.}\
  \bibnamefont {Wang}}, \bibinfo {author} {\bibfnamefont {N.}~\bibnamefont
  {Bonini}}, \bibinfo {author} {\bibfnamefont {N.}~\bibnamefont {Marzari}},
  \bibinfo {author} {\bibfnamefont {N.}~\bibnamefont {Pugno}}, \bibinfo
  {author} {\bibfnamefont {G.}~\bibnamefont {Savini}}, \bibinfo {author}
  {\bibfnamefont {A.}~\bibnamefont {Lombardo}}, \ and\ \bibinfo {author}
  {\bibfnamefont {A.~C.}\ \bibnamefont {Ferrari}},\ }\href@noop {} {\bibfield
  {journal} {\bibinfo  {journal} {Nature Materials}\ }\textbf {\bibinfo
  {volume} {11}},\ \bibinfo {pages} {294–300} (\bibinfo {year}
  {2012})}\BibitemShut {NoStop}%
\bibitem [{\citenamefont {Tan}\ \emph {et~al.}(2014)\citenamefont {Tan},
  \citenamefont {Wu}, \citenamefont {Han}, \citenamefont {Zhao},\ and\
  \citenamefont {Zhang}}]{TanPH2014prb}%
  \BibitemOpen
  \bibfield  {author} {\bibinfo {author} {\bibfnamefont {P.-H.}\ \bibnamefont
  {Tan}}, \bibinfo {author} {\bibfnamefont {J.-B.}\ \bibnamefont {Wu}},
  \bibinfo {author} {\bibfnamefont {W.-P.}\ \bibnamefont {Han}}, \bibinfo
  {author} {\bibfnamefont {W.-J.}\ \bibnamefont {Zhao}}, \ and\ \bibinfo
  {author} {\bibfnamefont {X.}~\bibnamefont {Zhang}},\ }\href@noop {}
  {\bibfield  {journal} {\bibinfo  {journal} {Physical Review B}\ }\textbf
  {\bibinfo {volume} {89}},\ \bibinfo {pages} {235404} (\bibinfo {year}
  {2014})}\BibitemShut {NoStop}%
\bibitem [{\citenamefont {Cong}\ and\ \citenamefont {Yu}(2014)}]{CongC2014nc}%
  \BibitemOpen
  \bibfield  {author} {\bibinfo {author} {\bibfnamefont {C.}~\bibnamefont
  {Cong}}\ and\ \bibinfo {author} {\bibfnamefont {T.}~\bibnamefont {Yu}},\
  }\href@noop {} {\bibfield  {journal} {\bibinfo  {journal} {Nature
  Communications}\ }\textbf {\bibinfo {volume} {5}},\ \bibinfo {pages} {4709}
  (\bibinfo {year} {2014})}\BibitemShut {NoStop}%
\bibitem [{\citenamefont {Jiang}\ \emph {et~al.}(2008)\citenamefont {Jiang},
  \citenamefont {Tang}, \citenamefont {Wang},\ and\ \citenamefont
  {Su}}]{JiangJW2008prb}%
  \BibitemOpen
  \bibfield  {author} {\bibinfo {author} {\bibfnamefont {J.-W.}\ \bibnamefont
  {Jiang}}, \bibinfo {author} {\bibfnamefont {H.}~\bibnamefont {Tang}},
  \bibinfo {author} {\bibfnamefont {B.-S.}\ \bibnamefont {Wang}}, \ and\
  \bibinfo {author} {\bibfnamefont {Z.-B.}\ \bibnamefont {Su}},\ }\href@noop {}
  {\bibfield  {journal} {\bibinfo  {journal} {Physical Review B}\ }\textbf
  {\bibinfo {volume} {77}},\ \bibinfo {pages} {235421} (\bibinfo {year}
  {2008})}\BibitemShut {NoStop}%
\bibitem [{\citenamefont {Lui}\ \emph {et~al.}(2012)\citenamefont {Lui},
  \citenamefont {Malard}, \citenamefont {Kim}, \citenamefont {Lantz},
  \citenamefont {Laverge}, \citenamefont {Saito},\ and\ \citenamefont
  {Heinz}}]{LuiCH}%
  \BibitemOpen
  \bibfield  {author} {\bibinfo {author} {\bibfnamefont {C.~H.}\ \bibnamefont
  {Lui}}, \bibinfo {author} {\bibfnamefont {L.~M.}\ \bibnamefont {Malard}},
  \bibinfo {author} {\bibfnamefont {S.}~\bibnamefont {Kim}}, \bibinfo {author}
  {\bibfnamefont {G.}~\bibnamefont {Lantz}}, \bibinfo {author} {\bibfnamefont
  {F.~E.}\ \bibnamefont {Laverge}}, \bibinfo {author} {\bibfnamefont
  {R.}~\bibnamefont {Saito}}, \ and\ \bibinfo {author} {\bibfnamefont {T.~F.}\
  \bibnamefont {Heinz}},\ }\href@noop {} {\bibfield  {journal} {\bibinfo
  {journal} {Nano Letters}\ }\textbf {\bibinfo {volume} {12}},\ \bibinfo
  {pages} {5539–5544} (\bibinfo {year} {2012})}\BibitemShut {NoStop}%
\bibitem [{\citenamefont {Plechinger}\ \emph {et~al.}(2012)\citenamefont
  {Plechinger}, \citenamefont {Heydrich}, \citenamefont {Eroms}, \citenamefont
  {Weiss}, \citenamefont {Schuller},\ and\ \citenamefont
  {Korn}}]{PlechingerG2012apl}%
  \BibitemOpen
  \bibfield  {author} {\bibinfo {author} {\bibfnamefont {G.}~\bibnamefont
  {Plechinger}}, \bibinfo {author} {\bibfnamefont {S.}~\bibnamefont
  {Heydrich}}, \bibinfo {author} {\bibfnamefont {J.}~\bibnamefont {Eroms}},
  \bibinfo {author} {\bibfnamefont {D.}~\bibnamefont {Weiss}}, \bibinfo
  {author} {\bibfnamefont {C.}~\bibnamefont {Schuller}}, \ and\ \bibinfo
  {author} {\bibfnamefont {T.}~\bibnamefont {Korn}},\ }\href@noop {} {\ \textbf
  {\bibinfo {volume} {101}},\ \bibinfo {pages} {101906} (\bibinfo {year}
  {2012})}\BibitemShut {NoStop}%
\bibitem [{\citenamefont {Zeng}\ \emph {et~al.}(2012)\citenamefont {Zeng},
  \citenamefont {Zhu}, \citenamefont {Liu}, \citenamefont {Fan}, \citenamefont
  {Cui},\ and\ \citenamefont {Zhang}}]{ZengHL}%
  \BibitemOpen
  \bibfield  {author} {\bibinfo {author} {\bibfnamefont {H.}~\bibnamefont
  {Zeng}}, \bibinfo {author} {\bibfnamefont {B.}~\bibnamefont {Zhu}}, \bibinfo
  {author} {\bibfnamefont {K.}~\bibnamefont {Liu}}, \bibinfo {author}
  {\bibfnamefont {J.}~\bibnamefont {Fan}}, \bibinfo {author} {\bibfnamefont
  {X.}~\bibnamefont {Cui}}, \ and\ \bibinfo {author} {\bibfnamefont {Q.~M.}\
  \bibnamefont {Zhang}},\ }\href@noop {} {\bibfield  {journal} {\bibinfo
  {journal} {Physical Review B}\ }\textbf {\bibinfo {volume} {86}},\ \bibinfo
  {pages} {241301} (\bibinfo {year} {2012})}\BibitemShut {NoStop}%
\bibitem [{\citenamefont {Zhao}\ \emph {et~al.}(2013)\citenamefont {Zhao},
  \citenamefont {Luo}, \citenamefont {Li}, \citenamefont {Zhang}, \citenamefont
  {Araujo}, \citenamefont {Gan}, \citenamefont {Wu}, \citenamefont {Zhang},
  \citenamefont {Quek}, \citenamefont {Dresselhaus},\ and\ \citenamefont
  {Xiong}}]{ZhaoYY}%
  \BibitemOpen
  \bibfield  {author} {\bibinfo {author} {\bibfnamefont {Y.}~\bibnamefont
  {Zhao}}, \bibinfo {author} {\bibfnamefont {X.}~\bibnamefont {Luo}}, \bibinfo
  {author} {\bibfnamefont {H.}~\bibnamefont {Li}}, \bibinfo {author}
  {\bibfnamefont {J.}~\bibnamefont {Zhang}}, \bibinfo {author} {\bibfnamefont
  {P.~T.}\ \bibnamefont {Araujo}}, \bibinfo {author} {\bibfnamefont {C.~K.}\
  \bibnamefont {Gan}}, \bibinfo {author} {\bibfnamefont {J.}~\bibnamefont
  {Wu}}, \bibinfo {author} {\bibfnamefont {H.}~\bibnamefont {Zhang}}, \bibinfo
  {author} {\bibfnamefont {S.~Y.}\ \bibnamefont {Quek}}, \bibinfo {author}
  {\bibfnamefont {M.~S.}\ \bibnamefont {Dresselhaus}}, \ and\ \bibinfo {author}
  {\bibfnamefont {Q.}~\bibnamefont {Xiong}},\ }\href@noop {} {\bibfield
  {journal} {\bibinfo  {journal} {Nano Letters}\ }\textbf {\bibinfo {volume}
  {13}},\ \bibinfo {pages} {1007} (\bibinfo {year} {2013})}\BibitemShut
  {NoStop}%
\bibitem [{\citenamefont {Zhang}\ \emph {et~al.}(2013)\citenamefont {Zhang},
  \citenamefont {Han}, \citenamefont {Wu}, \citenamefont {Milana},
  \citenamefont {Lu}, \citenamefont {Li}, \citenamefont {Ferrari},\ and\
  \citenamefont {Tan}}]{ZhangXprb2013}%
  \BibitemOpen
  \bibfield  {author} {\bibinfo {author} {\bibfnamefont {X.}~\bibnamefont
  {Zhang}}, \bibinfo {author} {\bibfnamefont {W.~P.}\ \bibnamefont {Han}},
  \bibinfo {author} {\bibfnamefont {J.~B.}\ \bibnamefont {Wu}}, \bibinfo
  {author} {\bibfnamefont {S.}~\bibnamefont {Milana}}, \bibinfo {author}
  {\bibfnamefont {Y.}~\bibnamefont {Lu}}, \bibinfo {author} {\bibfnamefont
  {Q.~Q.}\ \bibnamefont {Li}}, \bibinfo {author} {\bibfnamefont {A.~C.}\
  \bibnamefont {Ferrari}}, \ and\ \bibinfo {author} {\bibfnamefont {P.~H.}\
  \bibnamefont {Tan}},\ }\href@noop {} {\bibfield  {journal} {\bibinfo
  {journal} {Physical Review B}\ }\textbf {\bibinfo {volume} {87}},\ \bibinfo
  {pages} {115413} (\bibinfo {year} {2013})}\BibitemShut {NoStop}%
\bibitem [{\citenamefont {Liu}\ \emph {et~al.}(2014)\citenamefont {Liu},
  \citenamefont {Neal}, \citenamefont {Zhu}, \citenamefont {Tománek},\ and\
  \citenamefont {Ye}}]{LiuH2014}%
  \BibitemOpen
  \bibfield  {author} {\bibinfo {author} {\bibfnamefont {H.}~\bibnamefont
  {Liu}}, \bibinfo {author} {\bibfnamefont {A.~T.}\ \bibnamefont {Neal}},
  \bibinfo {author} {\bibfnamefont {Z.}~\bibnamefont {Zhu}}, \bibinfo {author}
  {\bibfnamefont {D.}~\bibnamefont {Tománek}}, \ and\ \bibinfo {author}
  {\bibfnamefont {P.~D.}\ \bibnamefont {Ye}},\ }\href@noop {} {\bibfield
  {journal} {\bibinfo  {journal} {ACS Nano}\ }\textbf {\bibinfo {volume} {8}},\
  \bibinfo {pages} {4033} (\bibinfo {year} {2014})}\BibitemShut {NoStop}%
\bibitem [{\citenamefont {Du}\ \emph {et~al.}(2010)\citenamefont {Du},
  \citenamefont {Ouyang}, \citenamefont {Shi},\ and\ \citenamefont
  {Lei}}]{DuY2010jap}%
  \BibitemOpen
  \bibfield  {author} {\bibinfo {author} {\bibfnamefont {Y.}~\bibnamefont
  {Du}}, \bibinfo {author} {\bibfnamefont {C.}~\bibnamefont {Ouyang}}, \bibinfo
  {author} {\bibfnamefont {S.}~\bibnamefont {Shi}}, \ and\ \bibinfo {author}
  {\bibfnamefont {M.}~\bibnamefont {Lei}},\ }\href@noop {} {\bibfield
  {journal} {\bibinfo  {journal} {Journal of Applied Physics}\ }\textbf
  {\bibinfo {volume} {107}},\ \bibinfo {pages} {093718} (\bibinfo {year}
  {2010})}\BibitemShut {NoStop}%
\bibitem [{\citenamefont {Fujii}\ \emph {et~al.}(1982)\citenamefont {Fujii},
  \citenamefont {Akahama}, \citenamefont {Endo}, \citenamefont {Narita},\ and\
  \citenamefont {Shirane}}]{FujiiY1982ssc}%
  \BibitemOpen
  \bibfield  {author} {\bibinfo {author} {\bibfnamefont {Y.}~\bibnamefont
  {Fujii}}, \bibinfo {author} {\bibfnamefont {Y.}~\bibnamefont {Akahama}},
  \bibinfo {author} {\bibfnamefont {S.}~\bibnamefont {Endo}}, \bibinfo {author}
  {\bibfnamefont {S.}~\bibnamefont {Narita}}, \ and\ \bibinfo {author}
  {\bibfnamefont {Y.~Y.~G.}\ \bibnamefont {Shirane}},\ }\href@noop {}
  {\bibfield  {journal} {\bibinfo  {journal} {Solid State Communications}\
  }\textbf {\bibinfo {volume} {44}},\ \bibinfo {pages} {579} (\bibinfo {year}
  {1982})}\BibitemShut {NoStop}%
\bibitem [{\citenamefont {Yamada}\ \emph {et~al.}(1984)\citenamefont {Yamada},
  \citenamefont {Fujii}, \citenamefont {Akahama}, \citenamefont {Endo},
  \citenamefont {Narita}, \citenamefont {Axe},\ and\ \citenamefont
  {McWhan}}]{YamadaY1984prb}%
  \BibitemOpen
  \bibfield  {author} {\bibinfo {author} {\bibfnamefont {Y.}~\bibnamefont
  {Yamada}}, \bibinfo {author} {\bibfnamefont {Y.}~\bibnamefont {Fujii}},
  \bibinfo {author} {\bibfnamefont {Y.}~\bibnamefont {Akahama}}, \bibinfo
  {author} {\bibfnamefont {S.}~\bibnamefont {Endo}}, \bibinfo {author}
  {\bibfnamefont {S.}~\bibnamefont {Narita}}, \bibinfo {author} {\bibfnamefont
  {J.~D.}\ \bibnamefont {Axe}}, \ and\ \bibinfo {author} {\bibfnamefont
  {D.~B.}\ \bibnamefont {McWhan}},\ }\href@noop {} {\bibfield  {journal}
  {\bibinfo  {journal} {Physical Review B}\ }\textbf {\bibinfo {volume} {30}},\
  \bibinfo {pages} {2410} (\bibinfo {year} {1984})}\BibitemShut {NoStop}%
\bibitem [{\citenamefont {Kaneta}, \citenamefont {Katayama-Yoshida},\ and\
  \citenamefont {Morita}(1982)}]{KanetaC1982ssc}%
  \BibitemOpen
  \bibfield  {author} {\bibinfo {author} {\bibfnamefont {C.}~\bibnamefont
  {Kaneta}}, \bibinfo {author} {\bibfnamefont {H.}~\bibnamefont
  {Katayama-Yoshida}}, \ and\ \bibinfo {author} {\bibfnamefont
  {A.}~\bibnamefont {Morita}},\ }\href@noop {} {\bibfield  {journal} {\bibinfo
  {journal} {Solid State Communications}\ }\textbf {\bibinfo {volume} {44}},\
  \bibinfo {pages} {613} (\bibinfo {year} {1982})}\BibitemShut {NoStop}%
\bibitem [{\citenamefont {Kaneta}, \citenamefont {Katayama-Yoshida},\ and\
  \citenamefont {Morita}(1986)}]{KanetaC1986jpsj}%
  \BibitemOpen
  \bibfield  {author} {\bibinfo {author} {\bibfnamefont {C.}~\bibnamefont
  {Kaneta}}, \bibinfo {author} {\bibfnamefont {H.}~\bibnamefont
  {Katayama-Yoshida}}, \ and\ \bibinfo {author} {\bibfnamefont
  {A.}~\bibnamefont {Morita}},\ }\href@noop {} {\bibfield  {journal} {\bibinfo
  {journal} {Journal of the Physical Society of Japan}\ }\textbf {\bibinfo
  {volume} {55}},\ \bibinfo {pages} {1213} (\bibinfo {year}
  {1986})}\BibitemShut {NoStop}%
\bibitem [{\citenamefont {Kaneta}\ and\ \citenamefont
  {Morita}(1986)}]{KanetaC1986jpsj2}%
  \BibitemOpen
  \bibfield  {author} {\bibinfo {author} {\bibfnamefont {C.}~\bibnamefont
  {Kaneta}}\ and\ \bibinfo {author} {\bibfnamefont {A.}~\bibnamefont
  {Morita}},\ }\href@noop {} {\bibfield  {journal} {\bibinfo  {journal}
  {Journal of the Physical Society of Japan}\ }\textbf {\bibinfo {volume}
  {55}},\ \bibinfo {pages} {1224} (\bibinfo {year} {1986})}\BibitemShut
  {NoStop}%
\bibitem [{\citenamefont {Jiang}, \citenamefont {Rabczuk},\ and\ \citenamefont
  {Park}(2014)}]{JiangJW2014bpsw}%
  \BibitemOpen
  \bibfield  {author} {\bibinfo {author} {\bibfnamefont {J.-W.}\ \bibnamefont
  {Jiang}}, \bibinfo {author} {\bibfnamefont {T.}~\bibnamefont {Rabczuk}}, \
  and\ \bibinfo {author} {\bibfnamefont {H.~S.}\ \bibnamefont {Park}},\
  }\href@noop {} {\bibfield  {journal} {\bibinfo  {journal} {Preprint at
  http://arxiv.org/abs/1409.5297}\ } (\bibinfo {year} {2014})}\BibitemShut
  {NoStop}%
\bibitem [{\citenamefont {Gale}(1997)}]{gulp}%
  \BibitemOpen
  \bibfield  {author} {\bibinfo {author} {\bibfnamefont {J.~D.}\ \bibnamefont
  {Gale}},\ }\href@noop {} {\bibfield  {journal} {\bibinfo  {journal} {J. Chem.
  Soc., Faraday Trans.}\ }\textbf {\bibinfo {volume} {93}},\ \bibinfo {pages}
  {629} (\bibinfo {year} {1997})}\BibitemShut {NoStop}%
\bibitem [{\citenamefont {Kokalj}(2003)}]{xcrysden}%
  \BibitemOpen
  \bibfield  {author} {\bibinfo {author} {\bibfnamefont {A.}~\bibnamefont
  {Kokalj}},\ }\href@noop {} {\bibfield  {journal} {\bibinfo  {journal}
  {Computational Materials Science}\ }\textbf {\bibinfo {volume} {28}},\
  \bibinfo {pages} {155} (\bibinfo {year} {2003})}\BibitemShut {NoStop}%
\bibitem [{\citenamefont {Slater}, \citenamefont {Koster},\ and\ \citenamefont
  {Wood}(1962)}]{SlaterJC1962pr}%
  \BibitemOpen
  \bibfield  {author} {\bibinfo {author} {\bibfnamefont {J.~C.}\ \bibnamefont
  {Slater}}, \bibinfo {author} {\bibfnamefont {G.~F.}\ \bibnamefont {Koster}},
  \ and\ \bibinfo {author} {\bibfnamefont {J.~H.}\ \bibnamefont {Wood}},\
  }\href@noop {} {\bibfield  {journal} {\bibinfo  {journal} {Physical Review}\
  }\textbf {\bibinfo {volume} {126}},\ \bibinfo {pages} {1307} (\bibinfo {year}
  {1962})}\BibitemShut {NoStop}%
\end{thebibliography}
%

\end{document}